\documentclass[10pt]{iopart} 
\usepackage{graphicx,xcolor}
\usepackage{amssymb} 

\usepackage{setstack}
\usepackage{gensymb}
\begin{document}

\title[Transverse magnetic field influence on wakefield in plasmas]{Transverse magnetic field influence on wakefield in plasmas}
\author{Sita Sundar$^1$ and Zhandos A. Moldabekov$^{2, 3}$}

\address{$^1$Department of Aerospace Engineering, Indian Institute of Technology Madras, Chennai - 600036, India\\
$^2$~Institute for Experimental and Theoretical Physics, Al-Farabi Kazakh National University, 71 Al-Farabi str., 050040 Almaty, Kazakhstan\\$^3$Institute of Applied Sciences and IT, 40-48 Shashkin Str., 050038 Almaty, Kazakhstan}
\ead{sitaucsd@gmail.com}
\vspace{10pt}
\begin{indented}
\item[]April 2019
\end{indented}
\pdfminorversion=4

\begin{abstract}

We present the results of an investigation of the wakefield around a stationary charged grain in an external magnetic field with non-zero transverse component with respect to the ion flow direction. The impact of the orientation of magnetic field on the wake behavior is assessed. In contrast to previously reported  significant  suppression of the wake oscillations due to longitudinal  magnetic field applied along flow, 
in the presence of transverse to flow magnetic field the wakefield exhibits a  long range recurrent oscillations. Extensive  investigation for a wide range of parameters reveal that in the sonic and supersonic regimes the wake has strong dependence on the direction of the magnetic field and exhibits sensitivity to even a meager deviation of magnetic field from the longitudinal orientation. The tool obtained with the study of impact of transverse component of magnetic field on the wake around grain in streaming ions can be used to 
potentially maneuver the grain-grain interaction  to achieve controlled grain dynamics.

\end{abstract}

%
%
%
\ioptwocol

\section{Introduction}
Wakefield in a plasma manifests itself in a variety of physical phenomena and has gained tremendous interest in the recent years~\cite{Morfill:RMP2009, Bonitz:Book2014, Das:POP2005}. To understand these physical phenomena exhibited by complex plasma, e.g. dust acoustic waves, wake features, solitary structures etc, a number of schemes has been devised.
One of the fundamental research problems complex plasma physicists have dealt with is to acquire the knowledge about controlled behavior of dusty plasma.  This controlled behavior of grain-plasma dynamics has often been achieved by applying external electric~\cite{CPPDonko, Bastykova, Donk_2017, Sundar:POP2017, Sundar:PRE2019} and magnetic fields ~\cite{Asan, Edward:POP2016, Joost:PPCF2015, Sundar:PRE2018,Kodanova:IEEE2019,Kaw:POP2002,Abdirakhmanov:IEEE2019}. 

Study of dusty plasmas in vertical magnetic field was studied first by Fujiyama et al.~\cite{Fujiyama:JSAP1994}. They investigated experimentally the transport of silicon particles by modulated magnetic field. Salimullah et al~\cite{Salimullah:PLA1996} explored the inequalities of charge and number densities of electron, ion, and dust particle and presented the low-frequency dust-lower-hybrid modes in a dusty plasma. Low frequency waves and magnetoacoustic modes in magnetized dusty plasmas were investigated in Ref.~\cite{Rao:JPP1995}. This was taken further by Sato et al.~\cite{Sato:POP2001}. They discussed the drastic effect of  magnetic field on levitating dust cloud shape control and rotation for strongly coupled dusty plasmas due to ion drag on fine particles. They also reported the dependence of rotation on plasma density variation. In another work, a different perspective for impact of longitudinal  magnetic field on complex plasmas was presented by Samsonov et.al.~\cite{Samsonov:NJP2003}. They  depicted the agglomeration and levitation of magnetic grains. They also proposed the possibility of magnetic field induced crystal formation which promoted the interest in inter-disciplinary scientific interaction. 
Contemporary work for the same regime was carried out by  Yaroshenko et.al.~\cite{Yaroshenko:NJP2003}. Their explanation for the mutual interaction was based on dipole theory and they delineated that field-aligned individual particle containing chains observed in experiments is due to the dipole short-range force \cite{Tskhakaya, PhysRevE.93.053204}.

With the progress in understanding of complex plasma dynamics and transport, invesigation of wake formation behind grain started concurrently~\cite{Ludwig:NJP2012,Joost:PPCF2015, Sundar:PRE2018}. Observation of wake oscillations behind grain due to ion focusing is now one of the fascinating features displayed by complex plasmas and it has gained the reputation of one of the most important phenomena observed in dusty plasma simulations. 
In recent works, impact of magnetic field aligned along 
flow on these wakes has been investigated in great detail~\cite{Joost:PPCF2015, Sundar:PRE2018, Ludwig:EPJD2018, Carstensen:PRL2012,Miloch:NJP2018}, but, surprisingly, as far as the impact on wake is concerned, the influence of magnetic field perpendicular to the flow has not received significant attention yet. Moreover, the wakefield in an external magnetic field with non-zero both longitudinal and transverse components has not been explored so far.    Therefore, it is the aim of this work to develop fundamental insight on the effect of  transverse component of an external magnetic field on the wakefield in complex plasmas.

Magnetic field influences the behavior of charged plasma particles, i.e., ions and electrons, and hence affects the overall dynamics of the system.   Dust particulates are heavy and it takes comparatively higher strength of magnetic field to make the grain magnetized.   It is pertinent to ask about the role of magnetic field on dusty plasma phenomena especially the exciting wake field features reported for the case of grains in moving ions in the sheath region.
Nambu et al.~\cite{Nambu:PRE2001} pioneered the study of grain in magnetized ion flow. Their main analytical result was damping of wake features for grain in flowing ions in the presence of magnetic field applied along the ion flow. They illustrated the reduction in ion overshielding around grain eventually resulting in suppression of wakefield. A decrease in the interaction force with increasing longitudinal magnetic field strength was also recently reported by Carstensen et al.~\cite{Carstensen:PRL2012}.





When  the magnetic field is directed along the flow  or along the direction of electric field, the dynamics of grain can be studied in a comparatively simple manner, as in the case of $\mathbf{E}\parallel\mathbf{B}$, the  Lorentz force related to streaming velocity vanishes. However, when the direction of $\mathbf{B}$ is perpendicular to the flow, $\mathbf{E}\times\mathbf{B}$ drift complicates the dynamics (with $\mathbf{E}$ being an external electric field).
We have applied the  magnetic field with non-zero transverse component to explore this very complicated dynamics  with ion flow direction along $\mathbf{z}$. We present here the detailed numerical exploration for three different flow regimes (subsonic, sonic, and supersonic) and for a wide range of magnetization.  Furthermore, we compare our results with the case of longitudinal magnetic field as well as the case without magnetic field. The investigation has been performed with Maxwellian ion distribution 
which should be changed to non-Maxwellian for pressures greater than $10~{\rm Pa}$ and other pertinent situations. 



The outline of the paper is as follows. In Sec.~\ref{s:2} we introduce the numerical simulation scheme used and present the description of the method. In Sec.~\ref{s:3} we present the results regarding the impact of the magnetic field with non-zero transverse component on the wakefield. We start, in Sec.~\ref{s:3.1}, by considering purely transverse magnetic field case (i.e., with zero longitudinal component). Then, in Sec.~\ref{s:3.2}, we explore the intermediate regime with non-zero transverse and longitudinal components of magnetic field induction.  Finally, we present a summary and conclusion in Sec.~IV .

\section{Simulation details and plasma parameters}\label{s:2}

\subsection{Particle-in-cell simulation} \label{s:2.1}
The  wake features around a grain were explored with \textsc{COPTIC} particle-in-cell simulations~\cite{Hutch:POP2011}. 
 The simulated system consists of a stationary charged grain in the presence of streaming ions under the influence of an external magnetic field.
  Besides providing comparatively accurate solutions for even non-linear regime, the importance of the code lies in the fact that it can be customized to resolve the particle in the near neighborhood of grain by orders of magnitude than elsewhere by imposing a non-uniform grid in the neighborhood of the grain. The simulation set-up is similar to the one considered in the recent works~\cite{Ludwig:EPJD2018,  Sundar:PRE2018, Sundar:POP2017, Sundar:PST2018} except that we have magnetic field component  perpendicular as well as parallel (few cases) to  the flow direction.
Further numerical detail and fundamentals can be gleaned from the paper with descriptions about \textsc{coptic}~\cite{Hutch:POP2011, Hutch:POP2013}. 

Besides finite-sized objects, \textsc{coptic} facilitates the incorporation of  point-charge grains also. The grain considered in the present work is point-charge and is solved using particle-particle particle-mesh ($P^3M$) technique.  
The ion dynamics in six-dimensional phase space in the presence of the self-consistent electric field $-\nabla \phi$,  an optional external force {\bf D}~\cite{Hutch:POP2013} (this extra force $\mathbf{D}$ is zero in our simulations for the {\it shifted Maxwellian distribution}) and an external magnetic field $\mathbf{B}$ is delineated by the equation

\begin{equation}
 {m_i} \frac{d \mathbf{v}}{dt} = q \left[- \nabla {\phi} + \mathbf{v}  \times \mathbf{B} \right]+ \mathbf{D},
\end{equation}
here $q$ denotes the ion charge. 


In order to solve the Poisson equation, a second-order accurate finite-difference-scheme equivalent to Shortley-Weller approximation  has been adopted. 
The solution of this Poisson equation, with given electron and ion number densities $n_e$ and  $n_i$ respectively, 

\begin{equation}
\nabla^2{\phi}=(e n_e-qn_i)/\epsilon_0,
\end{equation}
endows us with the resultant potential. 
The interpolation of the electric field and the solution of the Poisson equation  is incorporated with higher precision and convergence as it uses compact difference stencil which is specialized in dealing with arbitrary oblique boundaries for Cartesian mesh as well. 
On the outer mesh-edge, the potential gradient along $\hat{z}$ is set to zero. 
For the present work, implementation of the dynamics of the lighter electron species is governed by the Boltzmann description, 

\begin{equation}
n_e = n_{e \infty}\exp(e\phi/T_e).
\end{equation}
The value of the  Boltzmann constant is taken as unity for the present work. It is important to mention here that the Boltzmann approximation of electrons restrict us in capturing all kinetic effects. 
The present work is a trade-off between the effects we propose to look forward and the computational cost incurred by implementing the particle treatment of electrons. 
%
%


\begin{figure}
\includegraphics[scale=0.55, trim = 6.5cm 12.5cm 2.1cm 2.5cm, clip =true, angle=0]{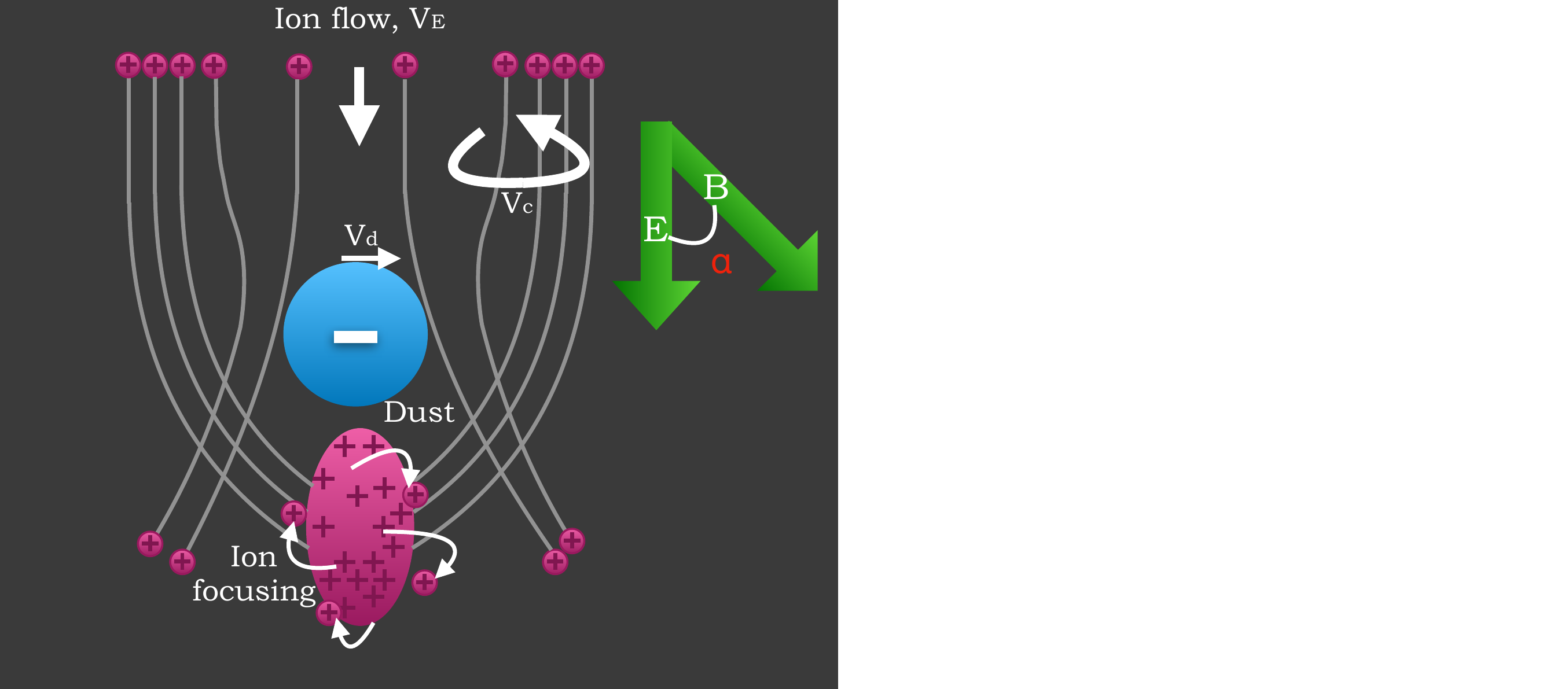}
\caption{Schematic depicting the system of grain in streaming ions in the presence of electric and magnetic fields.}
\label{fig:Figure1}
 \end{figure}

In the code, collisions are included according to the constant velocity-independent collision frequency  Poisson statistical distribution which is similar to the BGK-type collisions. Predominant collision in the system is the ion-neutral charged-exchange collision. This charge-exchange collision is implemented in the code by the mutual exchange of the ion velocity  with the neutral velocity randomly drawn  from neutral velocity distribution.

We considered a cell grid  of $64\times64\times128$ with more than 60 million ions and grid side length of $15\times15\times20$ Debye lengths for simulation purposes. We also performed few simulations with grids of even higher resolution and non-uniform mesh spacing to resolve the dynamics in the vicinity of the grain~\cite{Hutch:POP2011}. Simulation is progressed in time for 1000-2000 time steps (with units discussed below) by which it usually reaches steady-state.
 
\subsection{Dimensionless quantities and plasma parameters } \label{s:2.2} 

The most important \textit{dimensionless parameters} are as follows:
 \begin{enumerate}

\item The  orientation of the magnetic field 
is characterized by the angle $\alpha$ between  magnetic field induction vector and streaming (an external electric field) direction
(which is along $z$ axis) as illustrated in Fig.~\ref{fig:Figure1}. In general  we have $0\leq \alpha\leq \pi/2$. 
Clearly, the purely transverse magnetic field and purely longitudinal magnetic field cases correspond 
to $\alpha=\pi/2$ and $\alpha=0$, respectively. 
\item The strength of the magnetic field is conveniently and conventionally   characterized by the parameter $\beta=\omega_{\rm ci}/\omega_{\rm pi}$ defined as the ratio of cyclotron frequency of an ion to the plasma frequency of ions. Dimensionless parameters $\alpha$ and $\beta$ are sufficient to completely describe  the magnetic field. 

\item Mach number $M$ describing ionic streaming speed has been defined  as $M=v_d/c_s$, where $c_s=\sqrt{T_e/m_i}$ is the ion sound speed and $v_d$ is ionic streaming velocity. 
Thermal Mach number $M_{th}$ shares the relation with Mach number $M$ according to  the relation $M=v_d/c_s=\sqrt{T_i/T_e}\, M_{th}$.
\end{enumerate}

 

Besides, we follow the standard normalization, as described in~\cite{Hutch:POP2011}, i.e., the space coordinate is normalized as $r\rightarrow r/r_0$,  velocity as $v \rightarrow v/c_s $, and potential as $\phi \rightarrow \phi/(   T_e/e)$, where $r_0=( \lambda_{De}/5)$ is the normalizing scale length and $c_s$ is unity in normalized units. Time units as $\nu/(c_s/r_0) \sim 0.2(\nu/\omega_{pi})$ is used to normalize collision frequency $\nu$, where $\omega_{pi}$ is the ion plasma frequency. 
 The normalized grain charge, $Q_d$, is represented as ${\bar{Q_d}}={Q_d e}/({4\pi\epsilon_0 \lambda_{De} T_e})$,
 where $e_0$ is the unit electron charge.

 A summary of plasma parameters are presented in Table~\ref{table:TABLE I}.
  For typical dusty plasma parameters (e.g.  an electron Debye length of $\lambda_{de}= 845~{\rm \mu m}$ and an electron temperature of $2.585~{\rm eV}$), 
the normalized grain charge ${\bar{Q_d}}=0.01$ redacted in units of electron charge is approximately $7.5\times 10^3 e_0$ and $\beta\leq 0.35$ is the magnetization parameter corresponding to $\mathbf{B}<50~\mathrm{mT}$.


\begin{table}[h]
\caption{\label{arttype} List of plasma and simulation parameters.}
\footnotesize
\begin{tabular}{@{}llll}
\br
Physical property & Parameter range\\
\mr
\verb"Magnetization" $\beta$ & 0.0-1.0\\
\verb"Temperature ratio" $T_e/T_i$ & 100\\
\verb"Mach Number" $M$ & 0.5 - 1.5 \\
\verb"Collision frequency" $\nu/\omega_{pi}$  & 0.002\\
\verb"Normalized grain charge" $\bar{Q}_d$ & 0.01\\
\br
\label{table:TABLE I}
\end{tabular}\\
\end{table}
\normalsize

\section{Results}\label{sec:Res}\label{s:3}
Study of grain under streaming ions in the presence of cross-flow magnetic field 
serves the twofold objective of (a) investigation of the influence of magnetic field applied  perpendicular to the flow and (b) a  study of the intermediate case with no-zero transverse and longitudinal components of magnetic field.
In all presented figures a dust particle is located at the origin  $x=y=z=0$  (in Cartesian co-ordinate) and is post-processed to provide the result in cylindrical co-ordinate (with grain located at $z=0$ and $r=0$). 
\subsection{Impact of a transverse magnetic field  on the dust wake}\label{s:3.1}
 \begin{figure*}[htp]
 \includegraphics[scale=0.871, trim = 0.8cm 10.1cm 0.0cm 8.5cm, clip =true, angle=0]{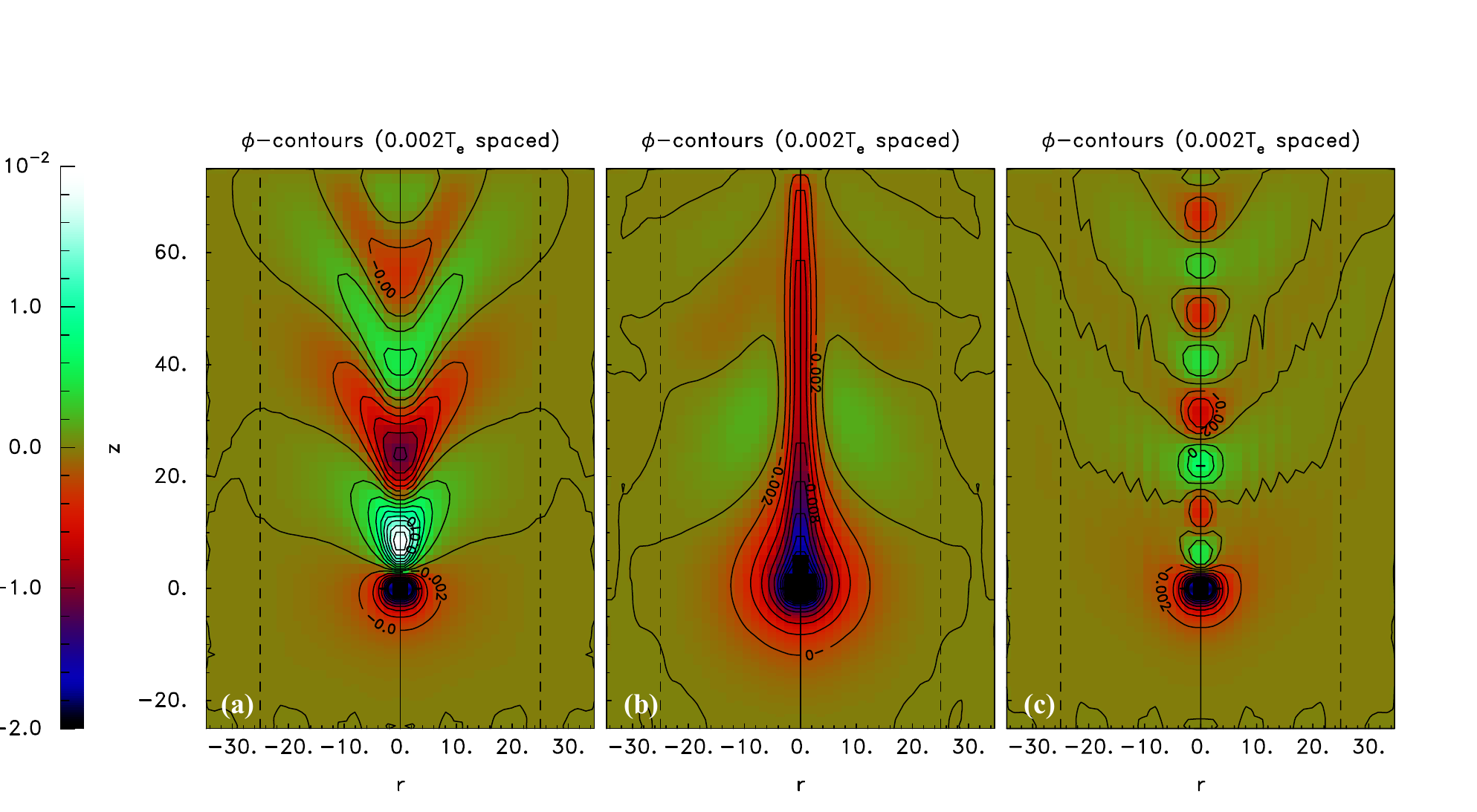} 
 \caption{\textcolor{black}{Wake potential contours $e \phi / T_e$, averaged over the azimuthal angle,  with the ion flow velocity $M=1.0$ 
 for: (a) without magnetic field ($\beta\sim 0.0$), (b) with magnetic field applied along streaming direction ($\beta=1.0$ and $\alpha=0$), and (c) with magnetic field in transverse direction  with respect to the flow($\beta=1.0$ and $\alpha=\pi/2$).}} 
\label{fig:Figure2}
 \end{figure*}
 
Let us start from the illustration of the wakefield at $M=1$ in an external magnetic field (with $\beta=1.0$) compared to that of  in the absence of  magnetic field (i.e. $\beta=0$). Accordingly, in Fig.~\ref{fig:Figure2} the contour plots of the potential in the cases:  (a) without magnetic field, (b) with magnetic field applied along  and (c) in transverse direction  with respect to the flow are shown.  Fig.~\ref{fig:Figure2} leads us to the first insight about the role played by the  magnetic field and its direction for the case of grain in streaming plasmas. We clearly see from Fig.~\ref{fig:Figure2} that the considered three cases have different pattern compared to each other. In the well studied magnetic field free  case, Fig.~\ref{fig:Figure2} (a), we see a  V-shaped wakefield with interchanging maxima and minima along flow direction. The wakefield with an external magnetic field applied along  streaming velocity---another in detail investigated situation---shows no oscillatory picture with ``candle flame" shape instead (see Fig.~\ref{fig:Figure2} (b)). The rotation of the magnetic field orientation from longitudinal to transverse direction restores oscillatory pattern of the wake field as illustrated in subplot (c) of Fig.~\ref{fig:Figure2}. The noticeable differences of the latter case from the magnetic field free case are a stronger localization of the ion focusing and depletion regions around maxima and minima accompanied by a weaker manifestation of the V-shape, and a weakened damping of these oscillations. We also notice the sustained long range oscillatory structures with reduced effective wake wavelength in the case of grain in transverse to flow magnetic field (cf. subplot (c)) and will be discussed later. In dimensional units, for typical dusty plasma parameters, e.g. $T_e=2.585$, $\beta=1.0$,  and $M=1$, the first wake peak would correspond to a potential of $60.62~{\rm mV}$ for magnetic field transverse to flow and $25.98~{\rm mV}$ for magnetic field along flow~\cite{Sundar:PRE2018}.

\begin{figure*}
\hspace{-1cm}
\includegraphics[ scale=1.175, trim = 2.5cm 5.95cm 0.0cm 6.1cm, clip =true, angle=0]{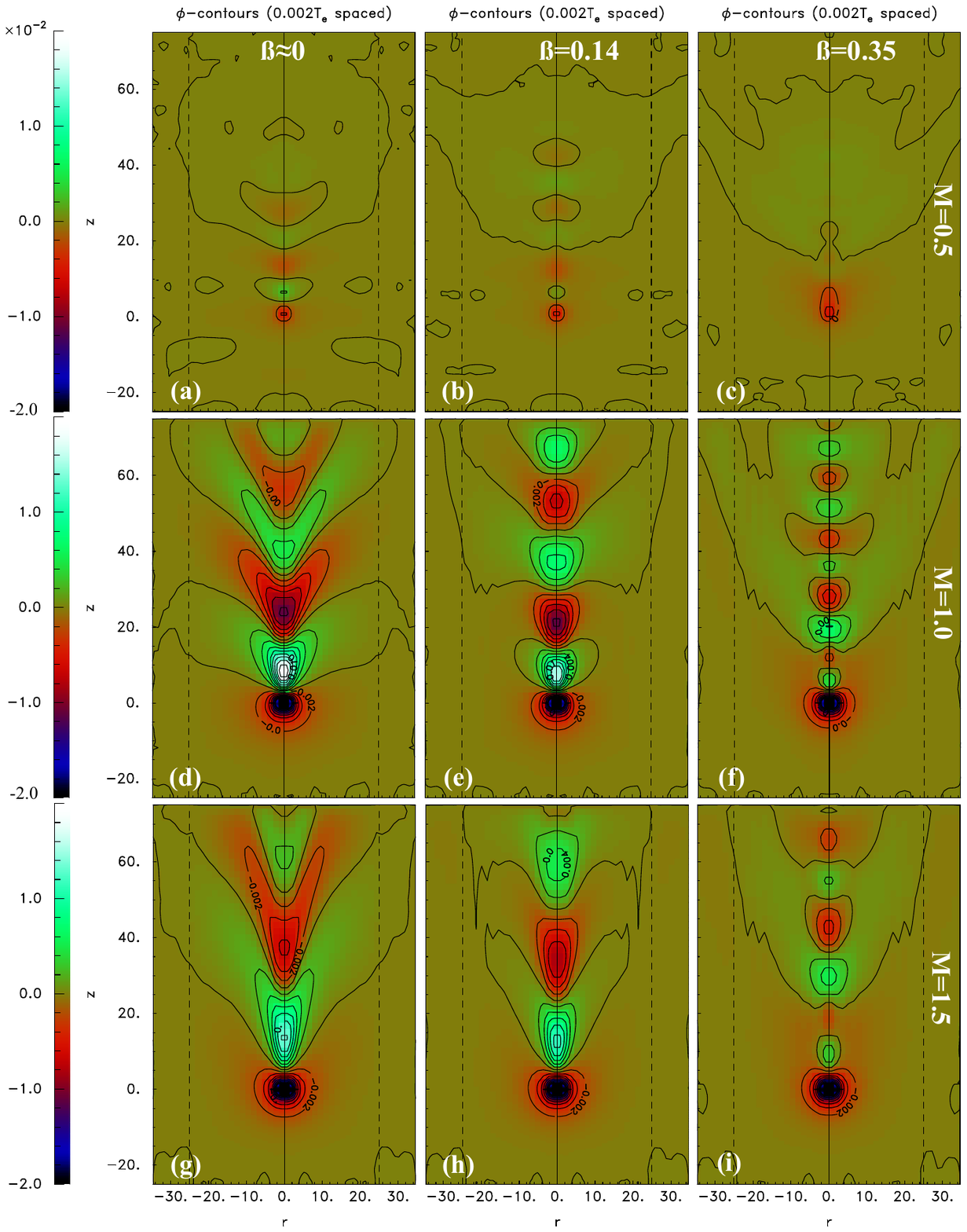} 
\caption{Wake potential contours $e \phi / T_e$ for various strengths of magnetic field.
The left column corresponds to  $\beta=0.07$, the middle column is for $\beta=0.14$, and the right column  is for $\beta=0.35$.
The ion flow velocity $M=0.5$ (top row), $M=1.0$ (middle row), and $M=1.5$ (bottom row).}
\label{fig:Figure3}
 \end{figure*}

Next, in Fig.~\ref{fig:Figure3}, we consider the wakefield in the transverse magnetic field with different magnetization parameter at subsonic, sonic, and supersonic cases.
 The three rows of the figure denote the cases with three different Mach numbers $M=0.5$, $M=1$, and $M=1.5$ (from top to bottom), while the three columns show the cases with increasing strengths of magnetic fields $\beta=0.07$, $\beta=0.14$, and $\beta=0.35$ (from left to right). For $M=0.5$ case (top row), magnetic field decreases the oscillations amplitude. The amplitude of wake oscillation is so small that no explicit trend is observed. Further increasing the magnetic field strength to $\beta=0.3535$,  we see an extended positive potential region at subsonic flows.  As we increased the flow speed to $M=1$ (middle row) and $M=1.5$ (bottom row), we see pronounced wake oscillations behind the grain. With the increase in $\beta$, the effect of magnetic field manifests in the increase of the number of oscillations and decrease in the distance between subsequent potential maxima and minima. Additionally,  the aforementioned stronger localization of the wake pattern around maxima and minima is distinctly visible.

 \begin{figure}
\hspace{-1cm}
\includegraphics[scale=0.55,trim = 1.0cm 1.0cm 0.0cm 0cm, clip =true,angle=0]{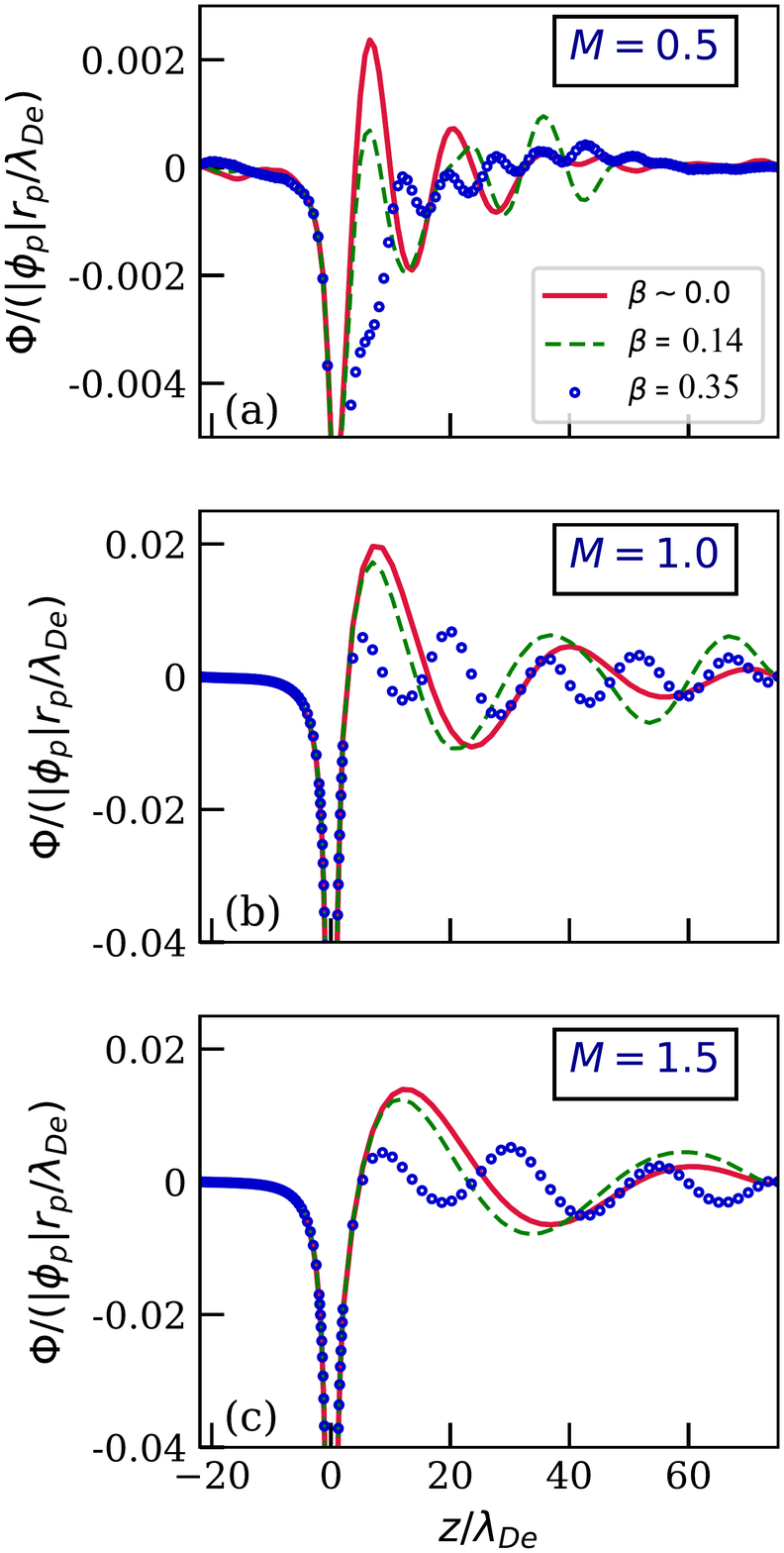}
\caption{Wake potential along the streaming axis for transverse to flow magnetic field ( $\alpha = \pi/2$) with $\beta\sim0.0$ (red solid line), $\beta=0.14$ (green dashed line), and $\beta=0.35$ (blue doted line) for $M=0.5$ (top panel), $M=1.0$ (middle panel), and $M=1.5$ (bottom panel).
}
\label{fig:Figure4}
\end{figure}

 \begin{figure}
\includegraphics[scale=0.41,trim = 0.5cm 8.75cm 0.0cm 6.5cm, clip =true,angle=0]{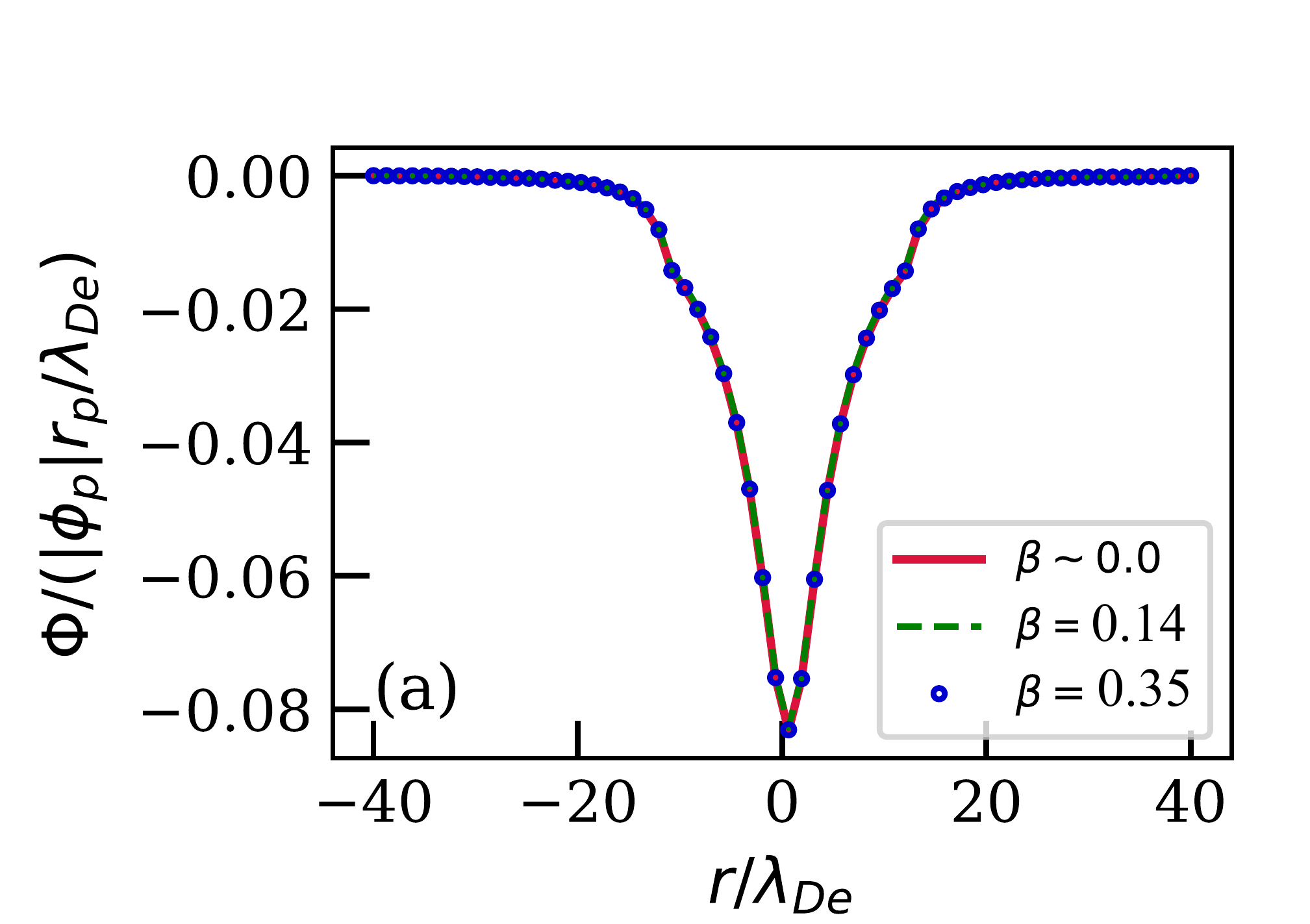}
\caption{Wake potential in transverse  to flow direction (with $z=0$, $\alpha = \pi/2$ and $M=1.0$) for magnetic field with $\beta=\sim 0$ (red solid line), $\beta=0.14$ (green dashed line), and $\beta=0.35$ (blue doted line).
}
\label{fig:Figure5}
\end{figure}

To better understand the impact of the transverse magnetic field,  we show in Fig.~\ref{fig:Figure4} the wake potential along the streaming axis for $M=0.5, 1$, and $1.5$, along with alomost unmagnetized or very feeble magnetic field case ($\beta=0.07$).  It was found that $\beta=0.07$ case provides almost the same data for the wake potential as the magnetic field free case already reported in previous works~\cite{Sundar:PRE2018, Sundar:POP2017, Joost:PPCF2015, Ludwig:NJP2012}. At subsonic and sonic regimes (top and middle panel, respectively),  the transverse magnetic field with $\beta=0.14$ leads to the deviation of the potential from the magnetic field free case. However, in the supersonic regime (bottom panel), data for $\beta=0.14$ case weakly differs from  $\beta=0.07\approx 0$ case. With further increase in the magnetic field strength to $\beta=0.35$, we see a strong impact of the transverse magnetic field in subsonic, sonic and supersonic regimes. In this case, we understand that the transverse magnetic field decreases the oscillations amplitude as well as the distance between maxima and minima of the wakefield. It is important to emphasize that while the oscillation amplitude is reduced, the damping of these  oscillations become weaker.

In Fig.~\ref{fig:Figure5}, the potential profile along transverse direction with $z=0$ is shown. This figure shows that the transverse magnetic field with $\beta\leq 0.35$ does not change the potential profile in  perpendicular direction to $z$ axis with $z=0$. This is an important information as in experiments often the dust particles  are located on a single plane perpendicular to the flow direction. Moreover, this behavior is in contrast to the longitudinal magnetic field case reported in previous works \cite{Sundar:PRE2018, Ludwig:EPJD2018, Joost:PPCF2015}, where it was shown that the longitudinal magnetic field has a strong impact on    the potential profile in  perpendicular direction to $z$ axis (streaming direction) with $z=0$.

\begin{figure}
\includegraphics[scale=0.41,trim = 0.5cm 8.75cm 0.0cm 6.5cm, clip =true,angle=0]{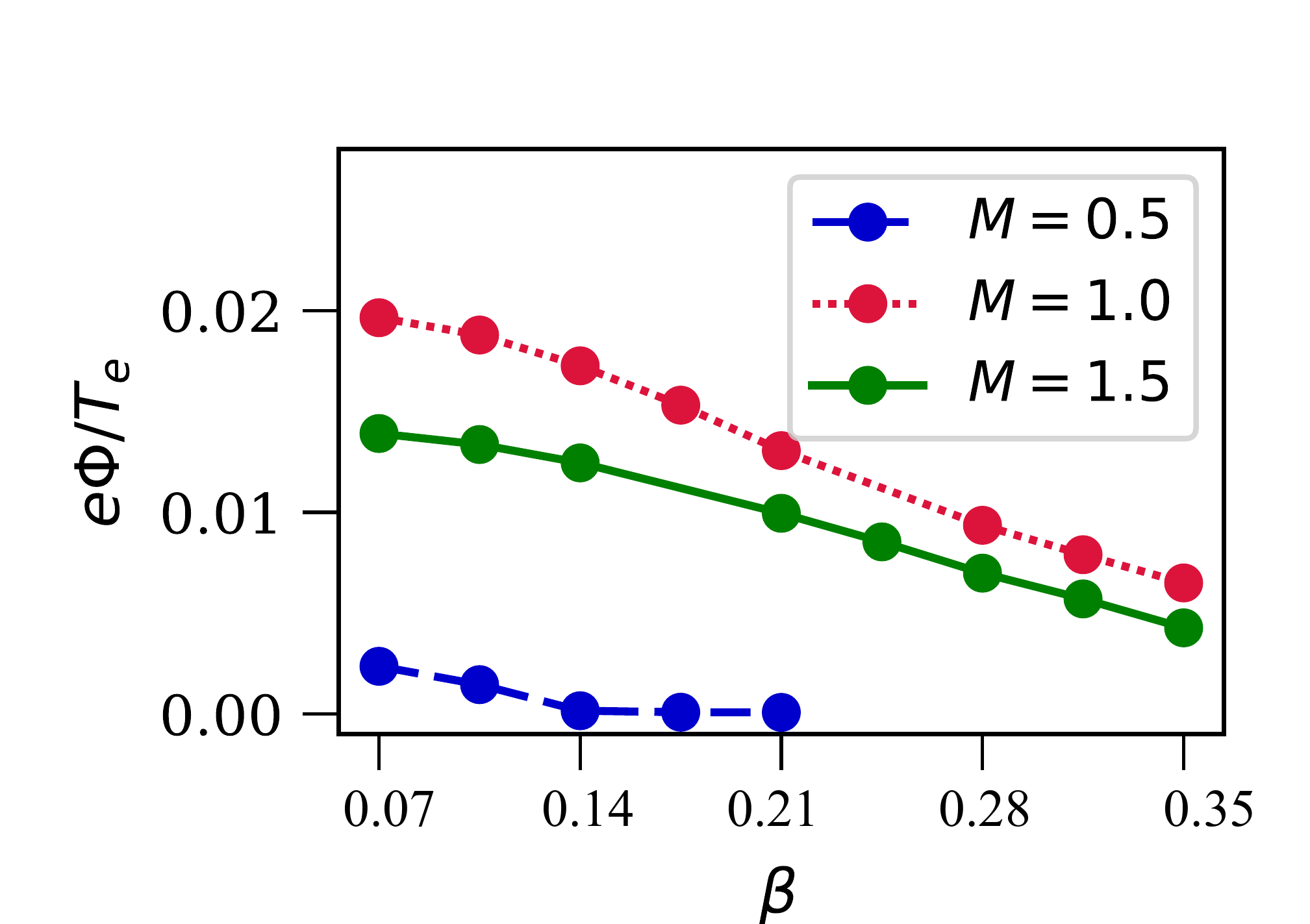}
\caption{Variation of the maximum of the 
peak amplitudes 
of the wake potential as a function
of $\beta$ with $\alpha = \pi/2$.
}
\label{fig:Figure6}
 \end{figure}

For completeness, for the case of transverse magnetic field, the  wake peak height  dependence on $\beta$ is given in Fig.~\ref{fig:Figure6} for $M=1.0$. 
In agreement with the above presented wake potential data, the wake peaks are substantially smaller for the case of $M=0.5$ compared to sonic and supersonic cases.  With increase in the flow speed, the wake peak amplitude increases ($M=1$) and then decreases ($M=1.5$). This behavior is due to the competition between increase in the number  of  influx ions and higher escape ability of ions with increase in streaming speed.  
Magnetic field applied perpendicular to the flow does not change this non-monotonic behavior with respect to change in the flow speed. The data for stronger magnetic field exhibits the smaller wake peak amplitude in coherence with that obtained for magnetic field applied along the flow~\cite{Sundar:PRE2018}. However, the relative damping of the oscillations is weaker in downstream direction compared to the longitudinal magnetic field case.




 \begin{figure*}
 \centering
 \includegraphics[width=0.75\textwidth]{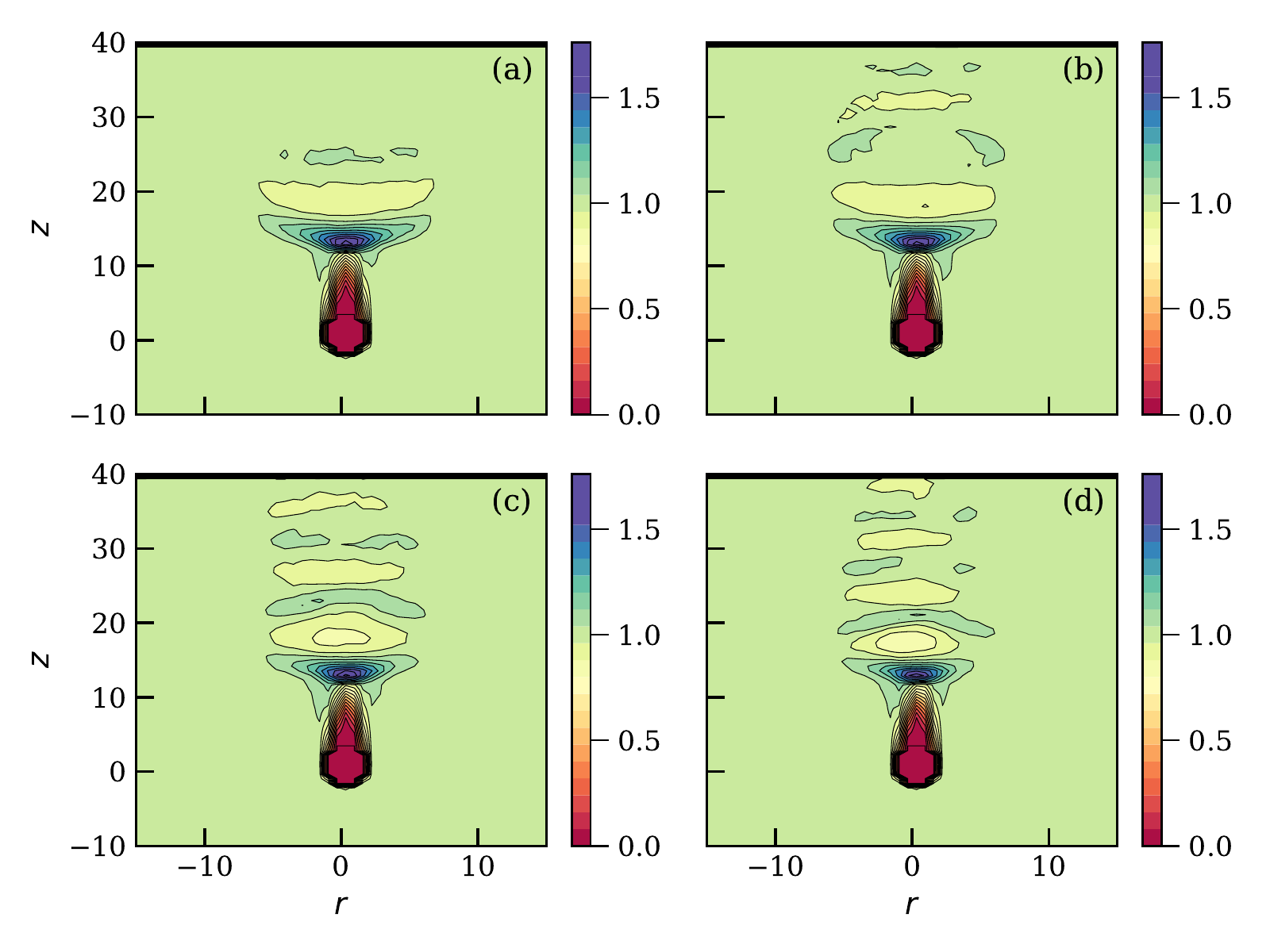}
\caption{
Spatial profiles of the ion density (normalized to the distant unperturbed ion density) for for $M=1.5$, $\alpha = \pi/2$, and  (a) $\beta=0.07$, (b) $\beta=0.14$, (c) $\beta=0.21$, and (d) $\beta=0.35$.
}
\label{fig:Figure7}
 \end{figure*}  
Noteworthy, the \textit{ion density} contour profile supports the discussed wakefield variation in the presence of transverse magnetic field as illustrated in Fig.~\ref{fig:Figure7} for flow speed $M=1.5$
for $\beta =0.07,0.14,0.21$, and $0.35$. We know from the work by Sundar {\it et al.}~\cite{Sundar:PRE2018} that the presence of longitudinal magnetic field induces the ion density fluctuation. Here, we see a magnetic field applied perpendicular to the flow direction also induces fluctuation in ion density, however, the fluctuation in the two cases exhibit a stark difference. In the case of magnetic field perpendicular to the flow, there is no `candle flame' like protruding structure in the density contour as obtained for the parallel to flow magnetic field case~\cite{Sundar:PRE2018}. Additionally, transverse magnetic field, makes the density fluctuation propagate farther from the grain. 

\subsection{Effect of deviation of magnetic field orientation from longitudinal direction}\label{s:3.2}

It is shown above that in the case of longitudinal magnetic field (i.e., when $ \mathbf{B}$ is directed along flow)  the wake behavior is completely different from that of  magnetic field applied perpendicular  to the flow. Therefore, it is interesting to inquire the intermediate case, i.e. when  both  longitudinal and transverse components  of the magnetic field are non-zero. The consideration of the intermediate configuration provides  picture about the influence of deviation of magnetic field orientation from the longitudinal direction on the wake features. 

\begin{figure*}
\includegraphics[scale=1.21, trim = 3.5cm 5.75cm 0.5cm 5.5cm, clip =true, angle=0]{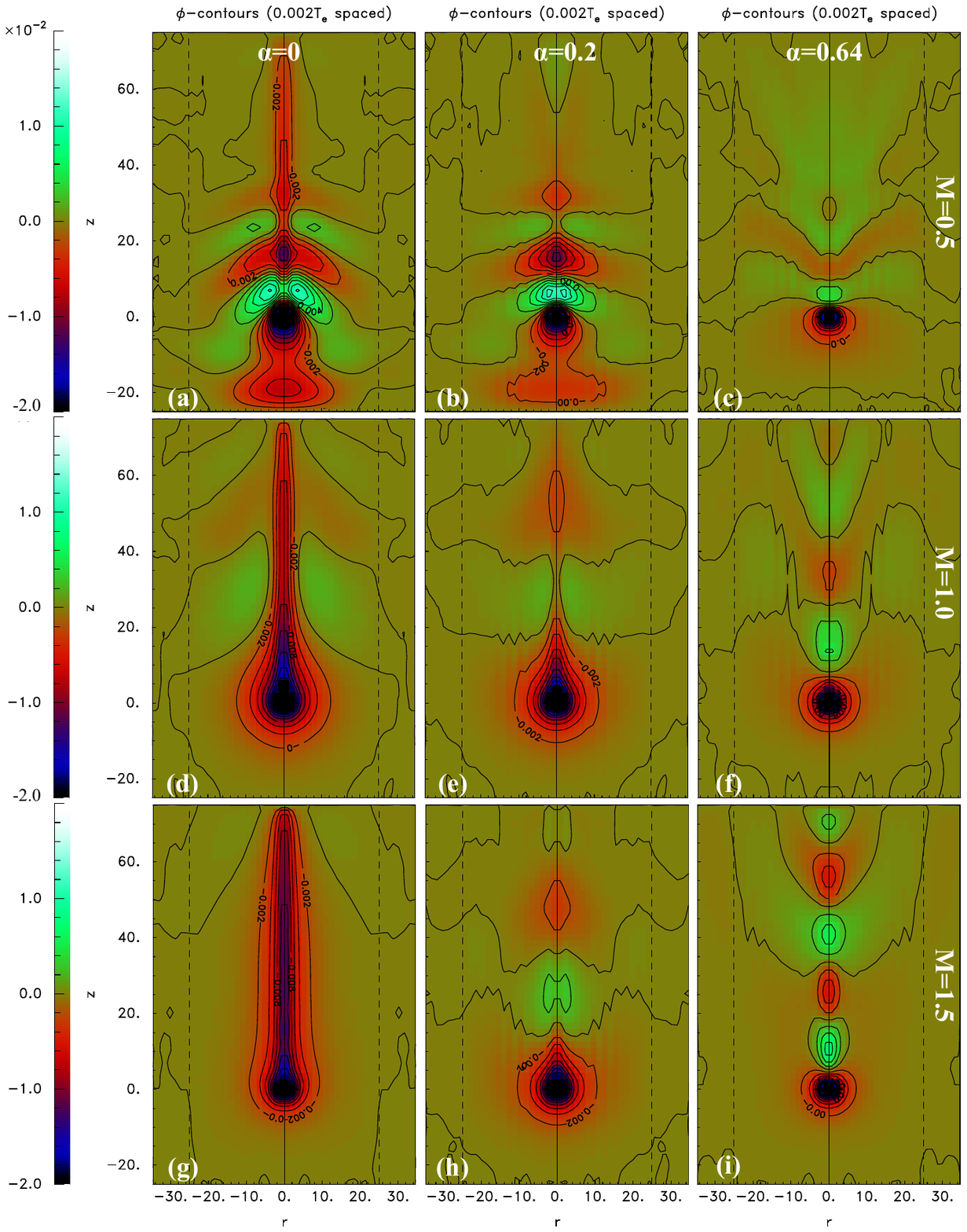} \\
\caption{Wake potential contours $e \phi / T_e$  with the ion flow velocity $M=1.0$ and magnetization parameter $\beta=1.0$  for various orientation angle of magnetic field $\alpha$,  where $\alpha$ is the angle between magnetic field direction and ionic streaming direction [i.e., $z$ axis]. From left to right column we have $\alpha=0$ [i.e. $\mathbf{B}=\mathbf{B_\parallel}$ with $ \mathbf{B_\perp}=0$ ], $\alpha=0.2$, and $\alpha=0.64$. 
}
\label{fig:Figure8}
 \end{figure*}

Figure~\ref{fig:Figure8}, depicts the impact of magnetic field orientation on the wake in the subsonic (top row), sonic (middle row), and supersonic (bottom row) cases. 
Here, we keep the strength of the overall magnetic field intact (with $\beta=1.0$) and vary $\mathbf{B_\parallel}$ and $ \mathbf{B_\perp}$ components.
In Fig.~\ref{fig:Figure8}, from left  to right,  the strength of the magnetic field  component $\mathbf{B_\parallel}$ (along  the ion flux) reduces and the strength of the component $\mathbf{B_\perp}$ increases. Accordingly, the angle between the ion flux direction and magnetic field induction direction, $\alpha$, changes from $0\degree$   [the left column]  to  $36\degree$ ($0.64~{\rm rad}$)[the right column]  with  intermediate value corresponding to $10\degree$ ($0.2~{\rm rad}$ ) [the middle column].  
The change in the orientation of magnetic field modifies the wake behavior substantially.
It is seen from  Fig.~\ref{fig:Figure8} that, in the sonic and supersonic cases, a small deviation of the magnetic field direction from the purely longitudinal case  has strong impact on the wakefield (compare the left pair of columns). This is especially strongly manifested in the supersonic case ($M=1.5$, see the bottom row), where $10\degree$ deviation of the magnetic field induction from the  longitudinal direction creates absolutely different pattern of the wake field in comparison with the purely longitudinal case (i.e., $\mathbf{B}=\mathbf{B_\parallel}$). In contrast to the sonic and supersonic cases, in the subsonic regime the wake field modifies smoothly with the rotation of the magnetic field induction direction from transverse to longitudinal configuration.

\section{Discussion and conclusion}
The present work provides a comprehensive picture of the influence of the transverse magnetic field on the grain wakefield and reveals the important role played by the orientation of the magnetic field. The investigation has been performed with Maxwellian ion distribution and the effect of  non-Maxwellian  ions  in transverse magnetic field is topic for  forthcoming studies. We considered weak and moderate magnetization of ions with $\beta\leq 0.35$ in subsonic, sonic, and supersonic regimes with Mach number $M\leq 1.5$.  

We revealed several new features of wakefield due to transverse magnetic field.
First of all, in stark contrast to the case of magnetic field along flow \cite{Sundar:PRE2018, Joost:PPCF2015},  under the influence of the transverse magnetic field  the number of wake oscillations increases in comparison to the magnetic field free case. Secondly, the damping of the amplitude of wakefield oscillations is not as strong as that observed for parallel to the flow magnetic field case. Another important result  we demonstrated herein is that  the deviation of the orientation of magnetic field induction  from the parallel to flow direction leads to a significant wakefield modification in downstream region. 

 Magnetic field influences the wake amplitude as well as overall structure of the grain plasma dynamics (see e.g. \cite{KonopkaPRE, Puttscher, ISI:000357689500089}). 
 Magnetic field along flow has been known to provide shape control of dust cloud. Wake formation has also been attributed to vertical alignment of paired grains. Essentially, $\mathbf{B}$ field could be of utmost importance for future research for dust shape control.  Therefore,  the presented  results  are relevant for accurate description of 
 the complex plasma experiments in weak to moderate external magnetic field \cite{ Abdirakhmanov:IEEE2019}.    
 
 Furthermore, our results can also be useful to understand better other phenomena involving the  motion of particles in cross fields, e.g. the dynamics of scramjet mixing and charge focusing with electric or magnetic fields. 

On a final note, we mention that  the work studied herein should  be  of interest  for open challenges in astrophysics, e.g.  the dynamics of trapped particles in Van Allen Belts \cite{Ukhorskiy2014}, 
the Saturn ring spokes formation \cite{Goertz} as well as other astrophysical phenomena wherein the motion of dust particles encounter magnetic field
and streaming plasmas. 



\section{Acknowledgments}
S. Sundar would like to acknowledges
 the support and hospitality of IIT Madras India. 
This work was supported by the 
DRDO project via project no. ASE1718144DRDOASAM.
Zh. Moldabekov thanks the funding from the Ministry 
of Education and Science of the Republic of Kazakhstan via the grant  BR05236730 
 ``Investigation of fundamental problems of Phys. Plasmas and plasma-like media''.  Our numerical simulations were performed at the HPC cluster of IIT Madras.

%

\newpage



\providecommand{\newblock}{}

\end{document}